\documentclass{epl}

\title{Preferential attachment of communities: the same principle, but a higher level}
\shorttitle{Preferential attachment of communities}
\author{P. Pollner\inst{1} \and G. Palla\inst{1,2} \and T. Vicsek\inst{1,2}}
\institute{
  \inst{1} Department of Biological Physics, E\"otv\"os University, P\'azm\'any P.\ stny.\ 1A, H-1117 Budapest, Hungary\\
  \inst{2} Biological Physics Research Group of HAS, E\"otv\"os University, P\'azm\'any P.\ stny.\ 1A, H-1117 Budapest, Hungary
}
\pacs{89.75.Hc}{Networks and genealogical trees}
\pacs{89.75.Fb}{Structures and organization in complex systems}
\pacs{89.75.Da}{Systems obeying scaling laws}

\newcommand{\bea}{\begin{eqnarray}}
\newcommand{\eea}{\end{eqnarray}}
\newcommand{\f}{\frac}

\usepackage{graphicx}
\usepackage{amssymb}

\begin{document}

\maketitle

\begin{abstract}

The graph of communities  is a network emerging above the level of 
individual nodes in the hierarchical organisation of a complex system. 
In this graph the nodes correspond
 to communities (highly interconnected subgraphs, also called modules
 or clusters), and the links refer to members shared by two communities.
Our analysis indicates that the development of this modular structure
 is driven by preferential attachment, in complete analogy with the
 growth of the underlying network of nodes. 
We study how the links between communities are born in a growing 
co-authorship network, and introduce a simple model for the 
dynamics of overlapping communities.

\end{abstract}

\section{Introduction}

A wide class of complex systems occurring from the level of cells to
society can be described in terms of networks capturing the intricate 
web of connections among the units they are made of. Graphs corresponding
 to these real networks exhibit unexpected non-trivial properties,
{\it e.g.}, new kinds of degree distributions, anomalous diameter,
spreading phenomena, clustering coefficient, and correlations
\cite{watts-strogatz,barabasi-albert,albert-revmod,dm-book,barrat}.
In recent years, there has been a quickly growing interest in 
the structural sub-units of complex networks, associated with
more highly interconnected parts
\cite{domany-prl,gn-pnas,zhou,newman-fast,al-parisi,huberman,spektral,potts,scott-book,pnas-suppl,everitt-book,knudsen-book,newman-europhys}.
These sets of nodes are usually called
clusters, communities, cohesive groups, or modules, with no widely accepted,
 unique definition.
Such building blocks (functionally related  proteins \cite{ravasz-science,spirin-pnas}, industrial sectors \cite{onnela-taxonomy},
groups of people \cite{scott-book,watts-dodds}, cooperative players
\cite{play1,play2}, {\it etc.}) can play a crucial role in forming the structural and functional
properties of the involved networks. On the other hand, the presence of
 communities in networks is a relevant and
informative signature of the {\it hierarchical nature} of complex systems
\cite{ravasz-science,vicsek-nature,self-sim-coms}. 

Typically, the communities in a complex system are not 
isolated from each other, instead, they have overlaps,
 {\it e.g.}, a protein can be part of more than one functional 
unit \cite{protein-complexes}, and people can be members in different social
groups at the same time \cite{wasserman}.
This observation naturally leads to the definition of the 
{\it community  graph}: a network representing the connections between the 
communities, with the nodes referring to communities and links corresponding 
to shared members between the communities.  Accordingly, the community degree
 $d^{\rm com}$ of a community is given by the number of other communities
 it overlaps with, and is equal to the degree of the corresponding node in 
the community graph. The studies of the relevant statistics describing
 the community graph ({\it i.e.}, the degree distribution, clustering {\it etc.}) 
 of real networks have just begun \cite{our-nature}. 
So far, in the networks investigated, the community degree distribution
 was shown to decay exponentially for low and as a power law for higher
 community degree values. This means that fat tailed degree distributions
 appear at two levels in the hierarchy of these systems: both at the 
 level of nodes (the underlying networks are scale free), and at the
 level of the communities as well.

 {\it Preferential attachment} is a key concept in the field of
 scale-free networks. In a wide range of graph models the basic mechanism
 behind the emerging power law degree distribution is that the new nodes attach
 to the old ones with probability proportional to their degree
\cite{barabasi-albert,albert-revmod,dm-book}.
 Furthermore,
 in earlier works the occurrence of preferential
 attachment was directly demonstrated in several real world networks 
 with scale free degree distribution\cite{tamas-mer,barab-mer,newman-mer}.
The observed fat tails in the degree distribution of the community graphs
indicate that the mechanism of preferential attachment could be present 
{\it at the level of communities} as well. Our aim in the present manuscript is
 to examine the attachment statistics of communities in order to
 clarify this question. Our investigations
 focus on the development of the communities in the growing 
 co-authorship network of the Los Alamos cond-mat e-print archive 
\cite{cond-mat}, in which the nodes correspond to authors, and 
two authors become linked if they publish an article together.
By studying the time evolution of this system,we investigate
the dynamics of the new community links.  
For example, 
 when a previously unlinked
community is attached to another one, what are the size and community degree 
 statistics of that other community? Another, 
 closely related issue addressed in this paper is the appearance of 
 new members in the communities. The size
 distribution of the communities was found to be a power-law in 
 the system to be investigated \cite{our-nature}.
 Thus it is natural to address questions such as: What happens when a node 
belonging to none of 
 the communities suddenly joins a community? What are the size and
 community degree statistics of the community chosen? 

\section{The communities}
In the present work we study the dynamics of the communities in
 the Los Alamos cond-mat e-print archive \cite{cond-mat}, in which
 an article with $n$ authors contributes with $(n-1)^{-1}$ to the
 weight of the links between every pairs of its authors. (The dataset contains
 altogether 30739 nodes and 136065 links). The communities
are extracted with the Clique Percolation Method (CPM) 
\cite{our-nature,our-prl} at each time step, using the CFinder
 package freely downloadable from \cite{CFinder}. (Each time step
 corresponds to one month, and the data set contained 143 time
 steps from February 1992 to April 2004). The communities obtained
 by the CPM correspond to $k$-clique percolation clusters in the network. 
 The $k$-cliques are complete subgraphs of size $k$ (in which each
 node is connected to every other nodes). A $k$-clique percolation cluster 
is a subgraph containing $k$-cliques
 that can all reach each other through chains of $k$-clique
 adjacency, where two $k$-cliques are said to be adjacent if they share
 $k-1$ nodes. The $k$-clique percolation clusters can be best visualised
 with the help of $k$-clique templates, that are objects isomorphic 
 to a complete graph of $k$ vertices. Such a template can be placed 
onto any $k$-clique in the graph, and rolled to an adjacent 
 $k$-clique by relocating one of its vertices and keeping its 
 other $k-1$ vertices fixed. Thus, the $k$-clique percolation 
 clusters ($k$-clique communities) of a graph are all those 
 subgraphs that can be fully  explored by rolling a $k$-clique template 
in them but cannot be left by this template.

The main advantages of this community definition are that it is not
 too restrictive, it is local, it is based on the density of the links
 and it allows overlaps between the communities: a node can be
 part of several $k$-clique percolation clusters at the same time. 
The number of communities a given node $i$
 belongs to shall be referred to as the membership number $m_i$ of the
 node from now on. 

 When applied to weighted networks (such as the present co-authorship
 network), the CPM method has two parameters: 
the $k$-clique size $k$, 
 and a weight threshold $\omega^*$ (links weaker than $\omega^*$ are ignored).
  The criterion used to  fix these parameters is based on finding a 
 community structure as highly structured as possible. In the present 
 paper we stick to the optimal parameter values found in earlier studies 
 of  the same co-authorship network \cite{our-nature}, 
 given by $k=6$ and $\omega^*=0.1$.

\section{Determining attachment probabilities}

The method presented below can be applied in general to any
empirical  study of an attachment process where the main goal is to 
 decide whether the attachment is uniform or preferential 
 with respect to a certain property 
({\it e.g.}, degree, size, {\it etc.}) of the
 attached objects ({\it e.g.}, nodes, communities {\it etc.}). 
 If the  process is uniform with respect to a property $\rho$, then
 objects with a given $\rho$ are chosen at a rate given by the
 distribution of $\rho$ amongst the available objects.
However, if the attachment mechanism prefers 
high (or low) $\rho$ values, then objects with high 
(or low) $\rho$ are chosen with a higher rate compared to the $\rho$ 
distribution of the available objects. 
To monitor this enhancement,
 one can construct the cumulative $\rho$ distribution $P_t(\rho)$ of the
available objects  at each time step $t$, 
together with the un-normalised cumulative 
$\rho$ distribution of the objects chosen by the process between $t$ and $t+1$,
 denoted by $w_{t\rightarrow t+1}(\rho)$. The value of 
$w_{t\rightarrow t+1}(\rho^*)$ at a given $\rho^*$ equals to the number of
 objects chosen in the process between $t$ and $t+1$, that had a 
 $\rho$ value larger than $\rho^*$ at $t$.  To detect deviations from
 uniform attachment, it is best to accumulate the ratio 
of $w_{t\rightarrow t+1}(\rho)$ and $P_t(\rho)$ during the time evolution to
 obtain
\bea
W(\rho)=\sum_{t=0}^{t_{\rm max}-1}\f{w_{t\rightarrow t+1}(\rho)}
{P_t(\rho)}.
\eea
If the attachment is uniform with respect to $\rho$, then $W(\rho)$
 becomes a flat function. However, if $W(\rho)$ is an increasing function, 
then the objects with large $\rho$ are favoured, if it is a decreasing
function, the objects with small $\rho$ are favoured in the attachment
 process. The advantage of this approach is that 
the rate-like variable $w_{t\rightarrow  t+1}(\rho)$ associated to the time step 
between $t$ and $t+1$ 
is always compared to the $P_t(\rho)$ distribution 
at $t$. Therefore 
$W(\rho)$ is able to indicate preference 
(or the absence of preference) even when  $P_t(\rho)$ is slowly changing in 
time (as in the case of the community degree in the co-authorship 
network under investigation).

\begin{figure}[h]
\centerline{\includegraphics[width=\textwidth]{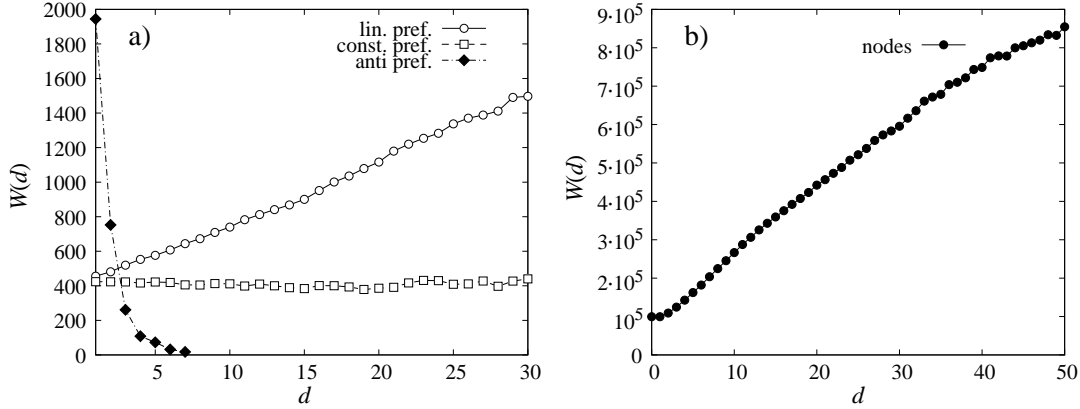}}
\caption{a) The  $W(d)$ function for networks grown 
 with known attachment rules: uniform probability (squares), 
linear preferential attachment (open circles), and anti preferential attachment 
(diamonds). b) The $W(d)$ function in the co-authorship network
 of the Los Alamos cond-mat archive.
\label{fig:test}}
\end{figure}

We have tested the above method on simulated graphs grown with known 
attachment mechanisms, {\it i)} uniform attachment (new nodes
 are attached to a randomly selected old node), {\it ii)} linear preferential
 attachment (new nodes are attached to old ones with a probability proportional
 to the degree), {\it iii)} and anti-preferential attachment 
(new nodes are attached to
 the old ones with a probability proportional to $\exp(-d)$, where $d$ is
 the degree). In these cases the degree $d$ of the individual nodes plays
 the role of the parameter $\rho$. For each time step, 
we recorded the cumulative 
degree distribution  of the nodes $P_t(d)$, together with 
 the number of nodes gaining new links with a degree higher than
 a given $d$, labelled by $w_{t\rightarrow t+1}(d)$. 
By summing the ratio of these two
 functions along the time evolution of the system one gets 
$W(d)=\sum_{t=0}^{t_{\rm max}-1}w_{t\rightarrow t+1}(d)/P_t(d)$.
In fig.\ref{fig:test}a. we show the empirical results for $W(d)$
 obtained for the simulated networks grown with the three 
different attachment rules.
The curves reflect the difference between the three cases very well:
for the uniform attachment probability $W(d)$ is flat, for
 the preferential attachment $W(d)$ is clearly increasing, and for 
 the anti-preferential attachment $W(d)$ is decreasing.  We have also
 calculated the attachment statistics of the nodes in the studied co-authorship
 network. As it can be seen in fig.\ref{fig:test}b.,  the corresponding 
$W(d)$ curve is increasing, therefore preferential attachment is
 present at the level of nodes in the system.

\section{Results}
In case of the communities of the investigated co-authorship network, 
 the two properties to be substituted in place of $\rho$ are the 
community degree $d^{\rm com}$
 and the community size $s$, therefore, the cumulative community size 
distribution $P_t(s)$ and the
 cumulative community degree distribution $P_t(d^{\rm com})$ were
 recorded at each time step $t$. To study the {\it establishment of the
 new community links}, we constructed the  un-normalised cumulative 
size distribution $w_{t\rightarrow t+1}(s)$ and the un-normalised 
cumulative degree 
 distribution $w_{t\rightarrow t+1}(d^{\rm com})$ of the communities 
gaining new community links to previously unlinked communities. 
 The
 value of these distributions at a 
given $s$ (or given $d^{\rm com}$)
 is equal to the number of
 unlinked communities at $t$ that establish a community link between
 $t$ and $t+1$ with a community larger than $s$ (or having larger degree than
 $d^{\rm com}$) at $t$. By accumulating the ratio of the rate-like
 variables and the corresponding distributions we obtain
\begin{equation}
W(s)=\sum_{t=0}^{t_{\rm max}-1} 
 \f{w_{t\rightarrow t+1}(s)}{P_t(s)}\, , \mbox{\hspace{2cm}}
W(d^{\rm com})=\sum_{t=0}^{t_{\rm max}-1}
\f{w_{t\rightarrow t+1}(d^{\rm com})}{P_t(d^{\rm com})}.
\end{equation}

For the investigation of the {\it appearance of new members} in the 
communities, we recorded the un-normalised community 
size distribution $\widehat{w}_{t\rightarrow t+1}(s)$ and the 
un-normalised community degree distribution 
$\widehat{w}_{t\rightarrow t+1}(d^{\rm com})$ of the 
communities gaining new members (belonging previously to none
 of the communities) between $t$ and $t+1$. The corresponding
 distributions that can be used to detect deviations from
 the uniform attachment are
\begin{equation}
\widehat{W}(s)=\sum_{t=0}^{t_{\rm max}-1} 
 \f{\widehat{w}_{t\rightarrow t+1}(s)}{P_t(s)}\, , \mbox{\hspace{2cm}}
\widehat{W}(d^{\rm com})=\sum_{t=0}^{t_{\rm max}-1}
\f{\widehat{w}_{t\rightarrow t+1}(d^{\rm com})}{P_t(d^{\rm com})}.
\end{equation}

\begin{figure}[h]
\centerline{\includegraphics[width=\textwidth]{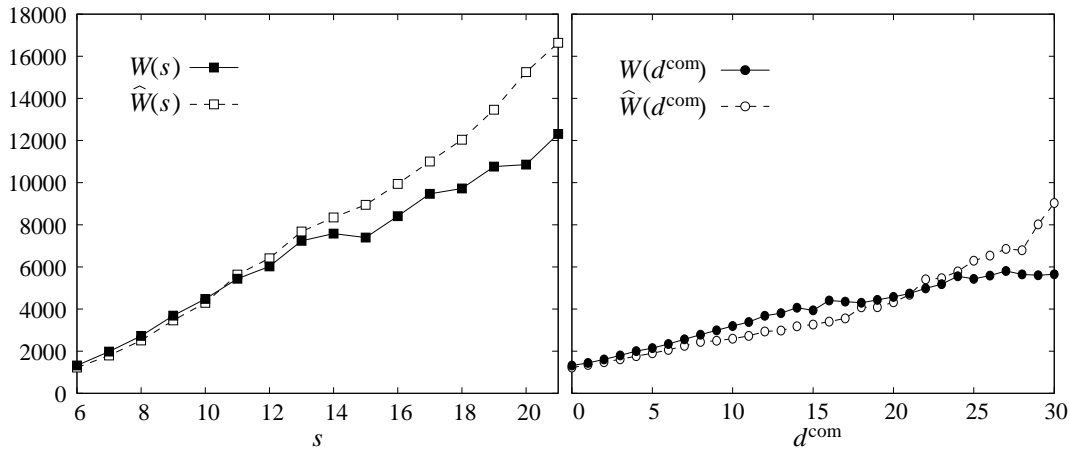}}
\caption{a) The $W(s)$ and $\widehat{W}(s)$ functions for the 
communities of the co-authorship network of the Los Alamos 
cond-mat e-print archive.
b) The $W(d^{\rm com})$ and $\widehat{W}(d^{\rm com})$ functions
 of the same network. The increasing nature of these functions indicates
 preferential attachment at the level of communities in the system.
\label{fig:com_results}}
\end{figure}

In fig.\ref{fig:com_results}a. we show the empirical $W(s)$ and
 $\widehat{W}(s)$ functions, whereas in fig.\ref{fig:com_results}b.
 the empirical $W(d^{\rm com})$ and $\widehat{W}(d^{\rm com})$ are
 displayed. All four functions are clearly increasing, therefore
 we can draw the following important conclusions:
\begin{itemize}
\item When a previously unlinked community establishes a new community
 link, communities with large size and large degree are selected with
 enhanced probability from the available other communities.
\item When a node previously belonging to none of the communities joins a  
community, communities with large size and large degree are selected with
 enhanced probability from the available communities.
\end{itemize}

We note that the community size and the community degree are strongly 
for higher sizes and degrees:
large communities have large community degree and vice versa.
Therefore, if one observes an attachment mechanism that is preferential
 with respect to either the community size, or the community degree, than
 it must be preferential for both of them.

\section{A toy model}

In this section we outline a simple model for the growth of overlapping
 communities, in which the preferential attachment of the node to 
 communities results in the emergence of a community system with
 scaling community size and community degree distribution. We note that when
 using the well known models based on preferential attachment 
solely between the nodes \cite{barabasi-albert,albert-revmod,dm-book},  
 the resulting graph  contains no communities at all at $k=6$.

In our model the underlying network between the nodes is left 
 unspecified, the focus is on the content of the communities.
During the time evolution, similarly to the models published in 
\cite{jeong-model,amaral-model,morris-model}, new members may join the already 
existing communities, and new communities may emerge as well. The new nodes
 introduced to the system choose their community preferentially with the
 community size, therefore the size distribution of the communities is
 expected to develop into a power-law. 
The appearance of the new community links originates in new nodes joining 
several communities at the same time.
 The detailed rules of the model are the  following:
\begin{itemize}
\item The initial state of the model is a small set of communities with random
 size.
\item The new nodes are added to the system separately. 
For each new node $i$, a membership $m_i$ is drawn from a Poissonean
 distribution with an expectation value of $\lambda$.
\item If $m_i\geq 1$, communities are succeedingly chosen with
 probabilities proportional to their sizes, until $m_i$ is reached, 
and the node $i$ joins the chosen communities simultaneously. 
\item If $m_i=0$, the node $i$ joins the group of unclassified vertices.
\item When the ratio $r$ of the group of unclassified nodes compared to the 
total number of nodes $N$ exceeds a certain limit $r^*$, 
a number of $q$ vertices from the group establish a new community.
(Obviously, $q$ must be smaller than $Nr$ even in the inital state). 
\end{itemize}
To be able to compare the results of the model with the community structure of
 the co-authorship network, the runs were stopped when 
 the number of nodes in the model reached the size of 
 the co-authorship network.

Our experience showed that the model
 is quite insensitive to changes in $r$ or $q$, and
 $\lambda$ is the only important parameter.  For small values ($\lambda<0.3$) 
 the resulting community degree distribution is truncated,
whereas when  $\lambda$ is too large
 ($\lambda>1$), a giant community with abnormally large community degree 
appears.
For intermediate $\lambda$ values ($0.3<\lambda<1$),
 the community size-- and community degree distributions become fat tailed,
 similarly to the co-authorship network.
\begin{figure}[h]
\centerline{\includegraphics[width=\textwidth]{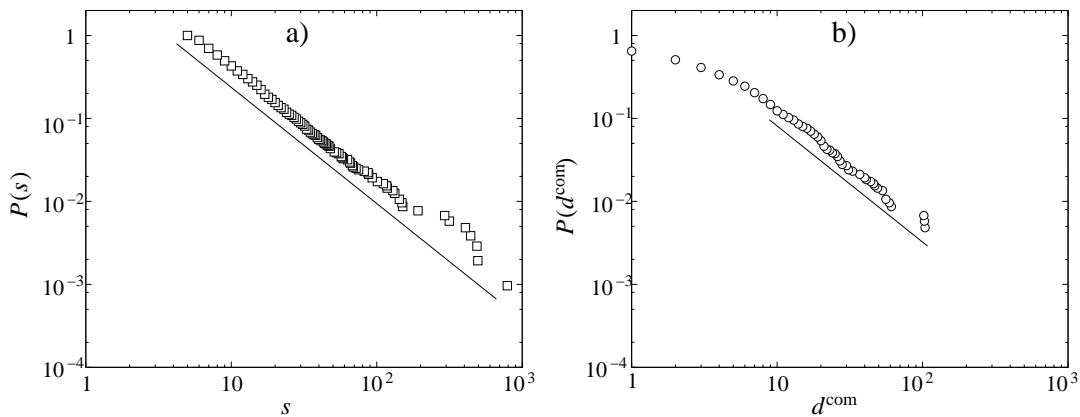}}
\caption{a) the cumulative community size distribution $P(s)$ (open circles) 
in our model at $\lambda=0.6$ follows a power-law with an exponent of $-1.4$
 (straight line) ( b) the cumulative community degree 
distribution $P(d^{\rm com})$ (filled circles) in our 
model at the same $\lambda$. The tail of this distribution follows the
 same power-law as the community size distribution (straight line), similarly
 to the communities found in the co-authorship network \cite{our-nature}.
\label{fig:toyresults}}
\end{figure}
 In fig.\ref{fig:toyresults}. we show the cumulative community size 
distribution $P(s)$ and the cumulative community degree distribution 
$P(d^{\rm com})$ of the communities obtained in our model at $\lambda=0.6$.
(Changes in the parameters $r$ and $q$ only shifts these curves,
 their shape remains unchanged).
 Our model grasps the relevant statistical properties of the 
community structure in the co-authorship network \cite{our-nature} quite well: 
 the community size distribution and the tail of the community degree 
distribution follow a power-law with the same exponent.

\section{Conclusions}

We studied the evolution of the community graph in a growing co-authorship
 network. We found that similar processes control the growth of the system at
 different levels in the hierarchy, as the growth of the communities, the 
 development of the community graph and the growth of the underlying network 
are all driven by preferential attachment. Inspired by these results,
we introduced a simple model for the dynamics of overlapping communities
 leading to scaling size-- and community degree distribution.

\acknowledgments
This work has been supported in part by
the Hungarian Science Foundation (OTKA), grant Nos. F047203 and T034995. 
 We thank A.-L. Barab\'asi for useful discussions.

\end{document}